\documentclass[prl,superscriptaddress,twocolumn]{revtex4}
\usepackage{enumerate}
\usepackage{amsfonts,amssymb,amsmath}
\usepackage[]{graphics,graphicx,epsfig}
\usepackage{amsthm}
\usepackage{float}
\usepackage{pdfpages}
\usepackage{epstopdf}
\DeclareGraphicsExtensions{.png,.pdf}
\usepackage{graphicx}
\usepackage{dcolumn}
\usepackage{natbib}
\usepackage{color}
\usepackage{multirow}
\usepackage{ulem}
\newcommand{\ket}[1]{|{#1}\rangle}
\newcommand{\bra}[1]{\langle{#1}|}

\usepackage{array}
\newcolumntype{P}[1]{>{\centering\arraybackslash}p{#1}}

\begin{document}

\title{In defense of temporal Tsirelson bound}
\author{Antoni W{\'o}jcik}
\affiliation{Institute of Spintronics and Quantum Information, Faculty of Physics, Adam Mickiewicz University, 61-614 Pozna\'n, Poland}

\author{Jan W{\'o}jcik}
\affiliation{Institute of Theoretical Physics and Astrophysics, University of Gdańsk, 80-308 Gda\'nsk, Poland}
\date{\today}

\begin{abstract}
In a recent paper, Chatterjee et al. [Phys. Rev. Lett 135, 220202 (2025)] analyze and experimentally implement a specific unitary evolution of a simple quantum system. The authors refer to this type of dynamics as a "superposition of unitary time evolutions." They claim that such an evolution enables a violation of the temporal Tsirelson bound in the Leggett-Garg scenario, a claim that is supported by their experimental results. In this work, we show that the proposed evolution can be understood within a more conventional framework, without invoking a superposition of evolutions. Furthermore, we demonstrate that the apparent violation of the bound arises because the measured quantities are not consistent with the necessary conditions for the temporal Tsirelson bound.
\end{abstract}

\maketitle
It is well established that a classical description of the physical world is incomplete. One of the most striking examples of this incompleteness is the limit imposed on correlations between measurement outcomes. Bell \cite{Bell1964} and Leggett-Garg \cite{Leggett1985} famously formulated inequalities that bound these correlations in space and time, respectively. Quantum mechanics, however, violates these bounds (for review see \cite{Brunner2014,Emary2014}), standing in stark contrast to the predictions of classical theory. These violations are not limitless and their bounds are usually called the Tsirelson bound and temporal Tsirelson bound for Bell and Leggett-Garg scenarios respectively.
 
Although temporal Tsirelson bound is well established within scientific community, in a recent paper, Chatterjee et al.\cite{Chatterjee2025} claim to experimentally violate it. Further the Authors claim that this violation is based on proposed new kind of quantum evolution. In this note we show to the contrary that not only the presented evolution is nothing more than a simple unitary transformation but also that the violation of temporal Tsirelson bound is only apparent.
 
To start with let us define operator $\mathcal{U}(t_f,t_0)$ as Eqs.(1) and (2) in \cite{Chatterjee2025} acting on 2-dimensional quantum systems. However, we will use slightly different notation $U_p(t)=\mathcal{U}(t_f,t_0)$,
 where $p=(\phi,\alpha)$, $t=\omega (t_f-t_0)/2$ using parameters ($\phi,\alpha,\omega,t_f,t_0$) introduced in \cite{Chatterjee2025}. Our notation emphasizes the fact that $\mathcal{U}(t_f,t_0)$ depends only on the difference $t_f-t_0$. Chatterjee et al.\cite{Chatterjee2025} construct $U_p(t)$ in the following way. First $\sigma_\phi$ is defined as a linear combination of Pauli operators
 \begin{equation}
     \sigma_\phi=\cos\phi \ \sigma_x+\sin\phi \ \sigma_y.
\end{equation}
 Then the time-dependent linear operator $W_p(t)$ is defined
\begin{equation}\label{sup}
     W_p(t)=\cos\alpha \ e^{-it\sigma_0}+\sin\alpha \ e^{-it\sigma_{\phi}}.
\end{equation}

After observation that $W_p^+(t)W_p(t)$ is proportional to identity $W_p^+(t)W_p(t)=n_p^2(t) I$, where
\begin{equation}
     n_p^2(t)=1+\left(cos^2(t)+\cos\phi \ sin^2(t) \right) sin(2 \alpha),
\end{equation}
 one constructs a unitary operator
\begin{equation}
     U_p(t)= W_p(t)/n_p(t).
\end{equation}
Explicit form of $W_p(t)$ is given by
\begin{widetext}
    \begin{equation}
     W_p(t)= \begin{pmatrix} cos(t)(\cos\alpha+\sin\alpha)&-isin(t)(\cos\alpha+e^{-i\phi)}\sin\alpha)\\-isin(t)(\cos\alpha+e^{i\phi)}\sin\alpha)&cos(t)(\cos\alpha+\sin\alpha)\end{pmatrix}.
    \end{equation}
\end{widetext}

The restriction $\phi \in [0,\pi)$ and $\alpha \in [0,\pi/2]$ guarantees the positivity of $n_p^2(t)$. 

 We believe that the specific form of Eq. \ref{sup} is the reason why $U_p(t)$ was referred to as a superposed unitary time evolution. In his comment, Gomez-Ruiz \cite{GomezRuiz2025} wrote, "Chatterjee and colleagues (,,,) let a qubit evolve under a coherent superposition of two distinct unitary transformations. Conceptually, the system follows two dynamical paths at once - extending the superposition principle from describing what a system is to describing how it evolves". He further noted that "By coherently combining different dynamical paths, the system explores an enlarged space of temporal correlations that ordinary evolution cannot reach". These statements suggest the presence of two distinct modes of evolution. 
   \begin{figure}[t]
\includegraphics[width=0.3\textwidth]{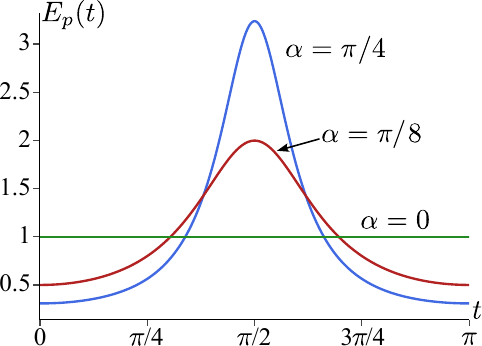}
\caption{\textbf{Time evolution of the Hamiltonian eigenvalue $\bm{E_p(t)}$.} 
    The plot shows the modulation of the energy eigenvalue $E_p(t)$ (Eq.~11) as a function of time $t$. The curves correspond to a fixed phase parameter $\phi=0.8$ and varying parameters: $\alpha=\pi/4$ (blue), $\alpha=\pi/8$ (red), and $\alpha=0$ (green). Note that $\alpha$ determines the amplitude of the modulation.}
\label{f1}
\end{figure}
Contrary to the above suggestion,  $U_p(t)$ 
is simply a family of unitary operators parametrized by $\phi,\alpha$, and $t$ and can therefore be used as a legitimate transformation of quantum states. Let us choose an arbitrary quantum state $\ket{\Psi_0}$. Acting with $U_p(t)$ generates a family of states
\begin{equation}
     \ket{\Psi_p(t)}=U_p (t) \ket{\Psi_0}.
\end{equation}
As $U_p(t=0)=I$ one has $\ket{\Psi_p(t=0)}=\ket{\Psi_0}.$
Dependence of $\ket{\Psi_p(t)}$ on parameter $t$ can be understood as an evolution of the quantum state with time $t$. The evolution is not uniform in the sense that it depends not only on the time difference but also on the initial state. This is because generally $U_p(t_1+t_2)\neq U_p(t_1)U_p(t_2)$.
It is convenient to introduce new unitary operator $\tilde{U}_p(t_2,t_1)$ such that for any $t_1, t_2$
\begin{equation}
    \ket{\Psi_p(t_2)}=\tilde{U}_p (t_2,t_1) \ket{\Psi_p(t_1}).
\end{equation}

This operator can be constructed as 
\begin{equation}
    \tilde{U}_p (t_2,t_1)=U_p(t_2) U_p^\dagger(t_1).
\end{equation}
Accordingly, the system evolves under $\tilde{U}_p (t_2,t_1)$ and $U_p(t)$ is just a special case of  $\tilde{U}_p (t_2,t_1)$, namely $U_p(t)=\tilde{U}_p(t,0)$. One can compute the time-dependent Hamiltonian $H_p(t)$ that generates this evolution. 
\begin{equation}
    H_p(t)=i \left(\frac{d}{dt} \tilde{U}_p (t,t_0) \right) \tilde{U}_p^\dagger(t,t_0).
\end{equation}
It takes the form
\begin{equation}\label{ham}
     H_p(t)=E_p(t) (\ket{\psi_{p,+}}\bra{\psi_{p,+}}-\ket{\psi_{p,-}}\bra{\psi_{p,-}}),
\end{equation}
where $\ket{\psi_{p,\pm}}$ are time-independent orthonormal eigenvectors  and the corresponding eigenvalues are $\pm E_p(t)$. Therefore, the evolution under consideration is equivalent to that of a spin-$\frac{1}{2}$ particle embedded in a magnetic field with a fixed direction and a time-dependent magnitude, without invoking any notion of superposed evolutions.

Explicit form of the eigenvalues and projectors on eigenspaces is given by

\begin{equation}\label{en}
     E_p(t)=\frac{(\cos\alpha + \sin\alpha)\sqrt{1+ \cos\phi \ \sin2\alpha}}{n_p^2(t)},
\end{equation}

\begin{equation}\label{pro}
     \ket{\psi_{p,\pm}}\bra{\psi_{p,\pm}}=\frac{1}{2}\begin{pmatrix} 1& \pm e^{-i \gamma}\\\pm e^{i \gamma}&1\end{pmatrix},
\end{equation}
where $\gamma$ fulfills
\begin{equation}
     tg \ \gamma=\frac{\sin\phi \ \sin\alpha}{\cos\alpha+\cos\phi \ \sin\alpha}.
\end{equation}

Note that the Hamiltonian is periodic because  $E_p(t+\pi)=E_p(t)$. An example of the time dependence of $E_p(t)$ is presented in Fig.1. The parameter $\alpha$
which in Ref. \cite{Chatterjee2025} is responsible for a magnitude of superposition, in our formulation directly determines the depth (amplitude) of Hamiltonian modulation.   For any $t_1$ and $t_2$ we have $[H_p(t_1), H_p(t_2)]=0$ so the evolution operator is given by 
\begin{equation}
    \tilde{U}_p(t_2,t_1)=e^{-i \int_{t_1}^{t_2} H_p(t')dt'}.
\end{equation}
Let us define $\tau(t)=\int E_p(t) dt$. From Eq. \ref{en} one obtains 
\begin{equation}
     \tau(t)=\arctan \left(\sqrt{\frac{1+\cos\phi \ \sin2 \alpha}{1+\sin2 \alpha} \ \tan(t)} \right)+\pi \left\lfloor \frac{t}{\pi}+\frac{1}{2}\right\rfloor.
\end{equation}

This allows us to write
\begin{equation}
     \tilde{U}_p(t_2,t_1)=e^{-i(\tau(t_2)-\tau(t_1) P },
\end{equation}
where projector $P=\ket{\psi_{p,+}}\bra{\psi_{p,+}}-\ket{\psi_{p,-}}\bra{\psi_{p,-}}$.
It follows that
\begin{equation}
     U_p(t)=e^{-i\tau(t) P }.
\end{equation}

\begin{figure}[b]
\includegraphics[width=0.3\textwidth]{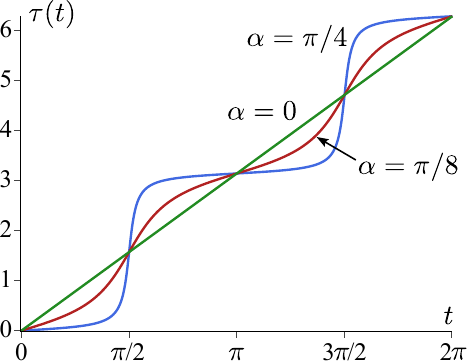}
\caption{\textbf{Dependence of the phase function $\tau(t)$ on time}. The three curves represent $\alpha=\pi/4$, $\alpha=\pi/8$, and $\alpha=0$ (with fixed $\phi = 0.95$).}
\label{f2}
\end{figure}

In Fig. \ref{f2} we present examples of function $\tau(t)$ for various parameters $\alpha$.

\section{Leggett-Garg scenario}

\begin{figure}[t]
\includegraphics[width=0.5\textwidth]{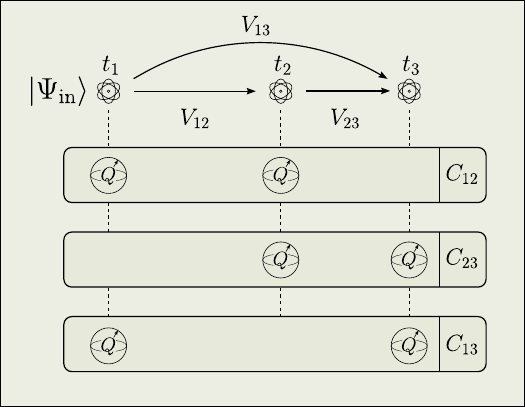}
\caption{\textbf{Schematic of the three-term Leggett-Garg scenario.} 
    The protocol involves three distinct experiments to measure temporal correlations $C_{jk} = \langle Q(t_j)Q(t_k) \rangle$ between fixed times $t_1 < t_2 < t_3$. In a valid scenario, the evolution between measurements must be consistent, satisfying $V_{13} = V_{23}V_{12}$.}
\label{f3}
\end{figure}

Let us now consider the three-term Leggett–Garg scenario. It consists of three distinct experiments (see Fig. \ref{f3}). All three share the following elements: three fixed times $t_1<t_2<t_3$, the same initial state prepared at $t=t_1$, and the same unitary evolution operators $V_{12}$ and $V_{23}$. In each experiment, the unitary evolution is interrupted at two distinct times by measurements of the same dichotomic observable $Q$. This allows one to estimate correlators $C_{12}$, $C_{23}$, and $C_{13}$, where $C{jk}=\left<Q(t_j)Q(t_k)\right>$. One can then compute the Leggett–Garg parameter $K_3=C_{12}+C_{23}-C_{13}$. The temporal Tsirelson bound is given by the inequality $K_3 \leq \frac{3}{2}$ \cite{Jordan2006,Kofler2008,Barbieri2009,Budroni2013}. Let us now concentrate on the specific case presented in \cite{Chatterjee2025}. Their initial state was a maximally mixed state $\frac{I}{2}$ and observable $Q=\sigma _z$. This allows us to write $C_{jk}=\frac{1}{2}tr\left[ \sigma _z V_{jk} \ \sigma _z V_{jk}^\dagger \right]$. A crucial requirement is that the evolution be identical in all three experiments. So, it must be
 \begin{equation}\label{cond}
     V_{13}=V_{23}V_{12}.
 \end{equation}

However, this condition is not satisfied in the analysis and experiment of Ref. \cite{Chatterjee2025}. Choosing $t_1=0,t_2=T$ and $t_3=2T$, Eq. (3) of \cite{Chatterjee2025} shows that the experimentally measured $K_3^{exp}$ is obtained using
$V_{12}=V_{23}=U_p(T)$ and $V_{13}=U_p(2T)$.
However, 
\begin{equation}
     U_p(2T) \neq U_p(T) \ U_p(T).
\end{equation}
Therefore, $K_3^{exp}$ cannot be regarded as a legitimate Leggett–Garg parameter. Consequently, one should not expect $K_3^{exp}$ to satisfy the bounds derived within the Leggett-Garg framework.

\begin{figure}[t]
\includegraphics[width=0.3\textwidth]{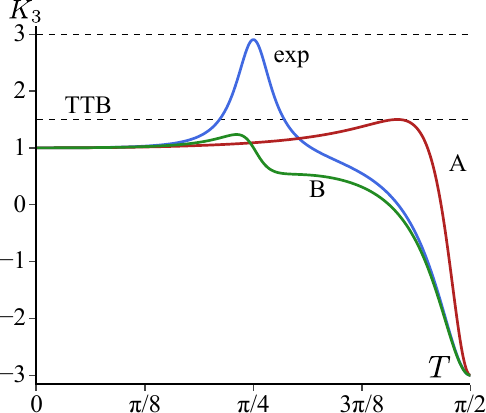}
\caption{\textbf{Leggett-Garg parameter $\bm{K_3}$ versus time interval $\bm{T}$.} 
    Comparison of $K_3$ for the original experiment (exp) and the two consistent scenarios (A and B) described in the text. The dashed line indicates the Temporal Tsirelson Bound (TTB = 1.5). Only the experimental curve (which violates the unitary composition law) exceeds the bound. Parameters used: $\phi=0.9\pi$ and $\alpha=\pi/4$. (Note: $T=\pi/2$ here corresponds to $t=\pi$ in \cite{Chatterjee2025})}
\label{f4}
\end{figure}

\begin{figure}[b]
\includegraphics[width=0.3\textwidth]{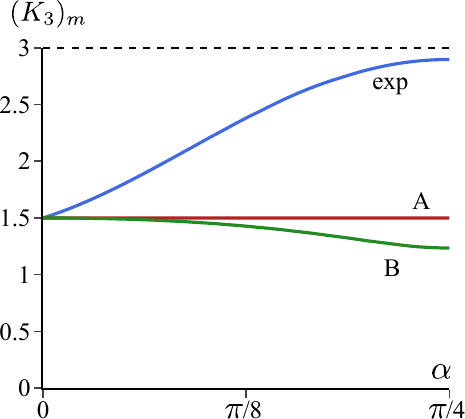}
\caption{\textbf{Maximal Leggett-Garg parameter $\bm{(K_3)_m}$ versus parameter $\bm{\alpha}$.} 
    The value of $K_3$ maximized over the time interval $T \in [0, \pi]$ is plotted against $\alpha$ (with fixed $\phi=0.9\pi$). The "exp" curve shows the apparent violation of the TTB (dashed line at 3) reported in [1]. The curves for the physically consistent scenarios A and B remain strictly within the bound $(K_3)_m \leq 1.5$.}
    \label{f5}
\end{figure}
\section{Modifications to the Experimental Protocol}

To address the inconsistency in the original experiment [1], where the unitary composition law ($V_{13} = V_{23}V_{12}$) was violated, we propose two modifications. These scenarios, labeled A and B, ensure a consistent definition of time evolution while utilizing the operators $U_p(t)$.

\begin{itemize}
    \item \textbf{Scenario A:} This scenario assumes a standard homogeneous evolution. We fix the single-step evolutions to be identical, $V_{12} = V_{23} = U_p(T)$. To satisfy the composition law, the total evolution operator is defined as the square of the single step: $V_{13} = U_p^2(T)$.
    
    \item \textbf{Scenario B:} This scenario accounts for the time-dependent nature of the underlying Hamiltonian. We define the evolution from $t_1=0$ to $t_2=T$ as $V_{12} = U_p(T)$, and the total evolution to $t_3=2T$ as $V_{13} = U_p(2T)$. Consequently, the intermediate step must be the correct propagator from $T$ to $2T$, given by $V_{23} = \tilde{U}_p(2T, T)$.
\end{itemize}

In both Scenarios A and B, the composition condition (Eq.~18) is strictly satisfied. Consequently, we find no violation of the Temporal Tsirelson Bound in either case. 

Figure~4 illustrates the dependence of the Leggett-Garg parameter $K_3$ on the time interval $T$. This data can be compared with Fig.~3f in \cite{Chatterjee2025} (noting the scaling difference: our $T = \pi/2$ corresponds to $t = \pi$ in their notation). 

Additionally, following \cite{Chatterjee2025}, we define $(K_3)_m$ as the maximum value of $K_3$ optimized over $T \in [0, \pi]$ for fixed $\phi$ and $\alpha$. Figure~5 presents $(K_3)_m$ as a function of $\alpha$ (with $\phi = 0.9\pi$). Comparison with Fig.~2e in \cite{Chatterjee2025} reveals that the violation of the bound disappears completely when the evolution operators are defined consistently.

In summary, we have addressed two primary issues. First, we examined the proposed concept of the superposition of unitaries. We demonstrated that the specific implementation used by Chatterjee et al. \cite{Chatterjee2025} simplifies to standard unitary dynamics governed by a time-dependent Hamiltonian. Consequently, we find this new terminology to be superfluous and potentially misleading \cite{GomezRuiz2025}, as it suggests the emergence of novel phenomena absent in the standard dynamics. In particular, Chatterjee et al. \cite{Chatterjee2025} claimed a violation of the temporal Tsirelson bound in the experimental realization of "superpositions of unitaries". 

Secondly, we found that the experiment fails to satisfy the necessary condition of temporal Tsirelson bound. We propose modifications that would render the experiment a valid temporal Tsirelson bound test and demonstrate that, in this case, the bound holds. Notably, these modifications are entirely feasible using the NMR techniques already developed by Chatterjee et al.

  \clearpage
\end{document}